\documentclass{article}

\PassOptionsToPackage{numbers, compress}{natbib}
\usepackage[preprint]{neurips_2020}

\PassOptionsToPackage{options}{natbib}
\usepackage{microtype}
\usepackage{graphicx}
\usepackage{booktabs} % for professional tables
\usepackage{amsmath}
\usepackage{amsthm}
\usepackage{subcaption}
\usepackage{wrapfig}
\usepackage{algorithm}
\usepackage{algorithmic}
\usepackage{amsmath}

\newtheorem{lemma}{Lemma}

\usepackage{verbatim}

\usepackage{cuted}
\usepackage{mathtools}
\usepackage{amssymb,amsfonts,latexsym,color}
\usepackage[utf8]{inputenc} % allow utf-8 input
\usepackage[T1]{fontenc}    % use 8-bit T1 fonts
\usepackage{hyperref}       % hyperlinks
\usepackage{url}            % simple URL typesetting
\usepackage{booktabs}       % professional-quality tables
\usepackage{amsfonts}       % blackboard math symbols
\usepackage{nicefrac}       % compact symbols for 1/2, etc.
\usepackage{microtype}      % microtypography
\usepackage{natbib}
\title{A Maximum Mutual Information Framework for Multi-Agent Reinforcement Learning}

\author{%
  Woojun Kim, Whiyoung Jung, Myungsik Cho, Youngchul Sung \\
  School of Electrical Engineering, KAIST, Korea\\
  \{woojun.kim, wy.jung, ms.cho, ycsung\}@kaist.ac.kr \\
}

\begin{document}

\maketitle

\begin{abstract}
In this paper, we propose a maximum mutual information (MMI) framework for multi-agent reinforcement learning (MARL) to enable multiple agents to learn coordinated behaviors by regularizing the accumulated return with the mutual information between actions. By introducing a latent variable to induce nonzero mutual information between actions and applying a variational bound, we derive a tractable lower bound on the considered MMI-regularized  objective function. Applying policy iteration to maximize the derived lower bound, we propose a practical algorithm named variational maximum mutual information multi-agent actor-critic (VM3-AC), which follows centralized learning with decentralized execution (CTDE). We evaluated VM3-AC for several games requiring coordination, and numerical results show that VM3-AC outperforms MADDPG and other MARL algorithms in multi-agent tasks requiring coordination.
\end{abstract}

%%%%%%%%%%%%%%%%%%%%%%%%%%%%%%%%%%%%%%%%%%%%%%%%%%%%%%%%%%%%%%%%%%%%%%%%%%%%%%
\section{Introduction}
\label{sec:Introduction}
%%%%%%%%%%%%%%%%%%%%%%%%%%%%%%%%%%%%%%%%%%%%%%%%%%%%%%%%%%%%%%%%%%%%%%%%%%%%%%including learning communication protocol \cite{2016foerester} and value-factorization \cite{2018rashid}

With the success of RL in the single-agent domain  \cite{2015Mnih,2015Lillicrap}, MARL is being  actively studied  and  applied to real-world problems such as  traffic control systems and connected self-driving cars, which can be  modeled as multi-agent systems requiring coordinated control
\cite{ 2019Li, 2019Andriotis}.
The simplest approach to MARL is independent learning, which trains each agent independently while treating other agents as a part of the environment.
One such example is independent Q-learning (IQL) \cite{1993Tan}, which is an extension of Q-learning to multi-agent setting. However, this approach suffers from the problem of non-stationarity of the environment.
A common solution to this  problem is to use  fully-centralized critic in the framework of centralized training with decentralized execution (CTDE) \cite{2019Oroojlooyjadid, 2018rashid}. For example, MADDPG \cite{2017Lowe} uses a centralized critic to train a decentralized policy for each agent, and COMA \cite{2018Foerster} uses a common centralized critic to train all decentralized policies.
However, these approaches assume that decentralized policies are independent and hence  the joint policy is the product of each agent's policy.
Such  non-correlated factorization of the joint policy limits the agents to learn coordinated behavior due to negligence of the influence of other agents \cite{2019Wen,2019de}.
However, learning coordinated behavior is one of the fundamental problems in MARL \cite{2019Wen, 2020Liu}.

In this paper, we introduce a new framework for MARL to learn coordinated behavior under CTDE without previously-used explicit dependency or communication in the execution phase.
Our framework is based on regularizing the expected cumulative reward with mutual information among agents' actions induced by injecting a latent variable.
The intuition behind the proposed framework is that agents can coordinate with other agents if they know what other agents will do with high probability, and the dependence between action policies can be captured by the mutual information.  High mutual information among actions means low uncertainty of other agents' actions.
Hence, by regularizing the objective of the expected cumulative reward with mutual information among agents' actions, we  can coordinate the behaviors of agents implicitly without explicit dependence enforcement.
However, the optimization problem with the proposed objective function has  several difficulties since we consider decentralized policies without explicit dependence or communication in the execution phase. In addition, optimizing mutual information is difficult because of the intractable conditional distribution.
We circumvent these difficulties by exploiting the property of the latent variable injected to induce mutual information, and
applying variational lower bound on the mutual information. With the proposed framework, we apply policy iteration by redefining value functions to propose the VM3-AC algorithm for MARL with coordinated behavior under CTDE. %The details will be explained in the upcoming sections.

Due to space limitation, related works are provided in Appendix A.

%%%%%%%%%%%%%%%%%%%%%%%%%%%%%%%%%%%%%%%%%%%%%%%%%%%%%%%%%%%%%%%%%%%%%%%%%%%%%%
\section{Background}
\label{sec:Background}
%%%%%%%%%%%%%%%%%%%%%%%%%%%%%%%%%%%%%%%%%%%%%%%%%%%%%%%%%%%%%%%%%%%%%%%%%%%%%%

We consider a Markov Game \cite{1994Littman}, which is an extention of Markov Decision Process (MDP) to multi-agent setting. An $N$-agent Markov game is defined by an environment state space $\mathcal{S}$, action spaces for $N$ agents $\mathcal{A}_1,\cdots,\mathcal{A}_N$, a state transition probability $\mathcal{T} : \mathcal{S} \times \boldsymbol{\mathcal{A}} \times \mathcal{S} \rightarrow [0,1]$, where $\boldsymbol{\mathcal{A}}=\prod_{i=1}^{N}\mathcal{A}_i$ is the joint action space, and a reward function $\mathcal{R}: \mathcal{S}\times \boldsymbol{\mathcal{A}} \rightarrow {\mathbb{R}}$. At each time step $t$, agent $i$ executes action $a_t^i\in \mathcal{A}_i$ based on  state $s_t\in \mathcal{S}$. The actions of all agents $\boldsymbol{a}_t=(a_t^1,\cdots,a_t^N)$ yields next state $s_{t+1}$ according to $\mathcal{T}$ and yields   shared common reward $r_t$ according to $\mathcal{R}$ under the assumption of fully-cooperative MARL.
%\tcr{or each agent can maximize local reward. }
The discounted return is defined as $R_t = \sum_{\tau=t}^{\infty} \gamma^{\tau} r_{\tau}$, where $\gamma \in [0,1]$ is the discounting factor.

We assume CTDE incorporating resource asymmetry between training and execution phases, widely considered in MARL \cite{2017Lowe,2018Iqbal,2018Foerster}.
%, and consider the partially observable setting where the agent $i$ can have access only to local observation $o_t^i$.
Under CTDE, each agent can access all information including the environment state, observations and actions of other agents in the system in the training phase, whereas the policy of each agent can be conditioned only on its own action-observation history $\tau_t^i$ or observation $o_t^i$ in the execution phase.
%For given joint policy $\boldsymbol{\pi}=(\pi^1,\cdots,\pi^N)$, the value functions for agent $i$ can be defined as $V_i^{\boldsymbol{\pi}}(s)=E_{\boldsymbol{\pi}}\big[R_0|s_0=s\big]$ and $Q_i^{\boldsymbol{\pi}}(s,\boldsymbol{a})=E_{\boldsymbol{\pi}}\big[R_0|s_0=s,\boldsymbol{a}_0=\boldsymbol{a} \big]$.
%Note that the state can be replaced with the observations of all agents $(o_1,\cdots,o_N)$, e.g. MADDPG used the centralized critic $Q_i^{\boldsymbol{\pi}}(o_1,\cdots,o_N, \boldsymbol{a})$.
    For given joint policy $\boldsymbol{\pi}=(\pi^1,\cdots,\pi^N)$, the goal of fully cooperative MARL is to find the optimal joint policy $\boldsymbol{\pi}^*$ that maximizes the objective $J(\boldsymbol{\pi})=E_{\boldsymbol{\pi}} \big[R_0\big]$.

\textbf{Maximum Entropy RL} The goal of maximum entropy RL is to find an optimal policy that maximizes the entropy-regularized objective function, given by
\begin{equation}\label{eq:MaxEntRL}
    J(\pi)=E_{\pi} \Bigg[\sum_{t=0}^{\infty} \gamma^t\Big( r_t(s_t,a_t)+\alpha H(\pi(\cdot|s_t))\Big)\Bigg]
\end{equation}
It is known that this objective encourages the policy to explore widely in the state and action spaces and helps the policy avoid converging to a local minimum.
Soft actor-critic (SAC), which is based on the maximum entropy RL principle, approximates soft policy iteration to the actor-critic method. SAC outperforms other deep RL algorithms in many continuous action tasks \cite{2018Haarnoja}.

We  can simply extend SAC to multi-agent setting in the manner of independent learning. Each agent trains decentralized policy using decentralized critic to maximize the weighted sum of the cumulative return and the entropy of its policy. We refer to this method as Independent SAC (I-SAC).
Adopting the framework of CTDE, we can replace decentralized critic with centralized critic which incorporates observations and actions of all agents. We refer to this method as multi-agent soft actor-critic (MA-SAC). Both I-SAC and MA-SAC are considered as baselines in the experiment section.

%%%%%%%%%%%%%%%%%%%%%%%%%%%%%%%%%%%%%%%%%%%%%%%%%%%%%%%%%%%%%%%%%%%%%%%%%%%%%%%
\section{The Proposed Maximum Mutual Information Framework}
\label{Method}
%%%%%%%%%%%%%%%%%%%%%%%%%%%%%%%%%%%%%%%%%%%%%%%%%%%%%%%%%%%%%%%%%%%%%%%%%%%%%%%

We assume that the environment is fully observable, i.e., each agent can observe the environment state $s_t$ for theoretical development in this section, and will consider the partially observable environment for practical algorithm construction under CTDE in the next section.

Under the proposed MMI framework, we aims to find the policy that maximizes the mutual information between actions in addition to cumulative return. Thus, the MMI-regularized objective function for joint policy $\boldsymbol{\pi}$ is given  by
\begin{equation}\label{eq:obj_mmi}
        J(\boldsymbol{\pi})=E_{\boldsymbol{\pi}} \Bigg[\sum_{t=0}^{\infty} \gamma^t\Big( r_t(s_t,\boldsymbol{a_t})+\alpha \sum_{(i,j)}I(\pi^i(\cdot|s_t);\pi^j(\cdot|s_t))\Big) \Bigg]
\end{equation}
where $a_t^i \sim \pi^i(\cdot|s_t)$ and $\alpha$ is the temperature parameter that controls the relative importance of the mutual information against the reward.

As aforementioned, we assume decentralized policies and want the decentralized policies to exhibit coordinated behavior. Furthermore, we want the coordinated behavior of the agents without explicit dependency previously used to enforce coordinated behavior.
Here, explicit dependency \cite{2019Jaques} means that for two agents $i$ and $j$, the action $a_t^i$ of agent $i$ follows $a_t^i \sim \pi^i(a^i_t|s_t)$ and then the  action $a_t^j$ of agent $j$ follows $a_t^j \sim \pi^j(a^j_t|s_t,a^i_t)$, i.e., the input to the policy function of agent $j$ explicitly requires the information about the action of agent $i$ for coordinated behavior.
By regularization with mutual information in the proposed objective function (\ref{eq:obj_mmi}), the policy of each agent is implicitly encouraged to coordinate with other agents' policies without explicit dependency by reducing the uncertainty about other agents' policies.
This can be seen as follows:
Mutual information is expressed in terms of the entropy and the conditional entropy as
\begin{equation}  \label{eq:MIform}
I(\pi^i(\cdot|s_t);\pi^j(\cdot|s_t))=H(\pi^j(\cdot|s_t))-H(\pi^j(\cdot|s_t)|\pi^i(\cdot|s_t)).
\end{equation}
If the knowledge of $\pi^i(\cdot|s_t)$ does not provide any information about $\pi^j(\cdot|s_t)$,
the conditional entropy reduces to the unconditional entropy, i.e.,
$H(\pi^j(\cdot|s_t)|\pi^i(\cdot|s_t)) = H(\pi^j(\cdot|s_t))$, and
the mutual information becomes zero.
 Maximizing mutual information is equivalent to minimizing the uncertainty about other agents' policies conditioned on the agent's own policy, which can lead the agent to learn coordinated behavior based on the reduced uncertainty about other agents' policies.
%Furthermore, by increasing the dependency between agents' actions by regularizing mutual information, the variance of learning  based on %stochastic gradient can be reduced.  Suppose that mutual information among agents' policies is maximized. Then, the maximum mutual %information case is the case that all other agents' actions are deterministic when one agent's action is given. Hence, the expectation in %\eqref{eq:obj_mmi} is now over a single variable not over $N$ variables, and the stochasticity reduces and the variance of learning can be %reduced.  Indeed, in MARL, it is known that the variance of learning increases as the number of agents increases \cite{2017Lowe}.
\begin{wrapfigure}{r}{6.5cm}
    \begin{tabular}{@{}cc@{}}
    \includegraphics[width=0.2\textwidth]{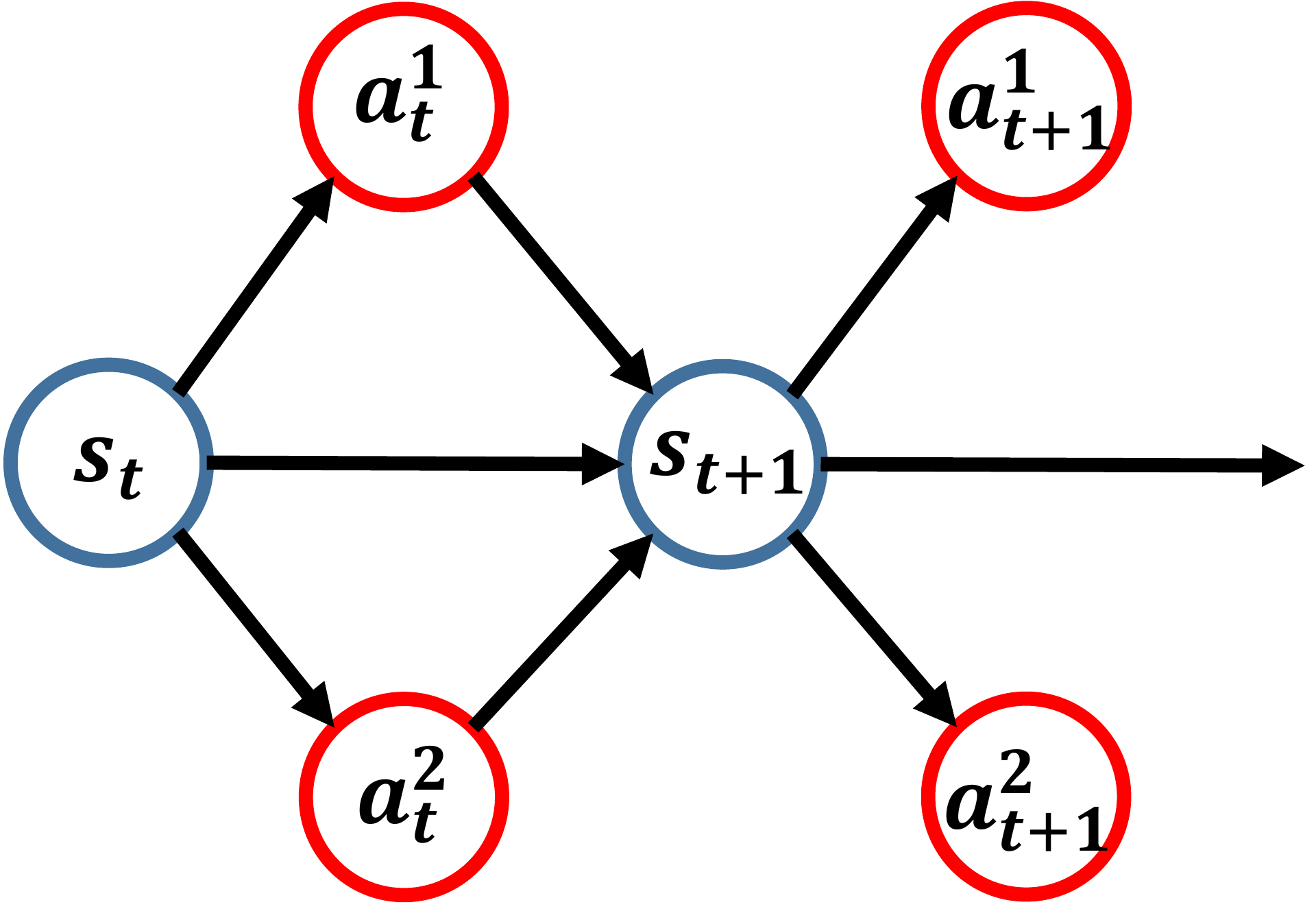}  &% Dummy image replacement
    \includegraphics[width=0.2\textwidth]{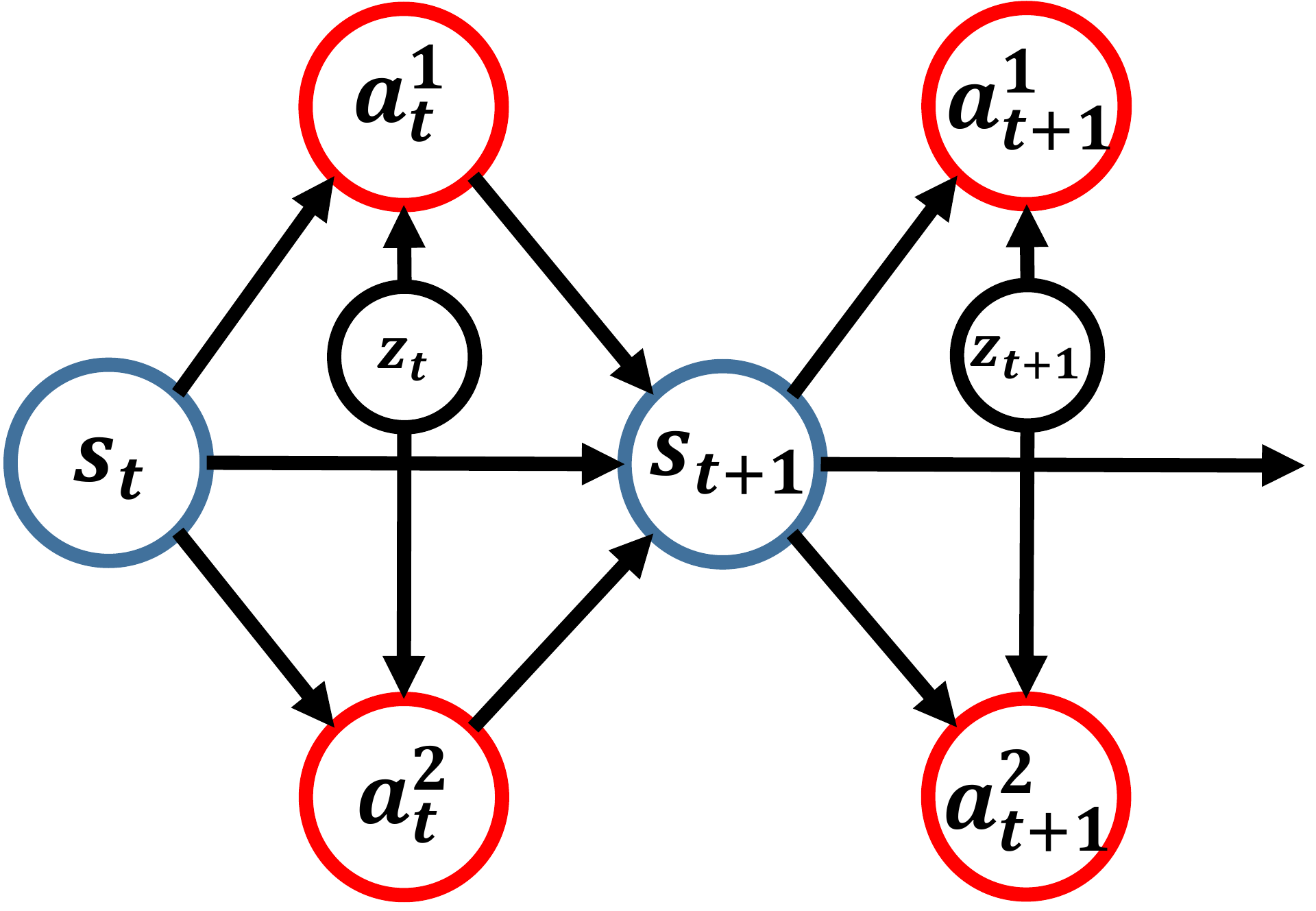}  \\% Dummy image replacement
    \end{tabular}
    \caption{Causal diagram in 2-agent Markov Game: (a) Standard MARL, (b) Introducing the latent variable to the standard MARL}
    \label{fig:twoagents}
\end{wrapfigure}
However, direct optimization of the objective function \eqref{eq:obj_mmi} is not easy.
Fig. \ref{fig:twoagents}(a) shows  the causal diagram of the  considered system model described in Section  \ref{sec:Background} in the  case of two agents with decentralized policies.  Since we consider the case of no explicit dependency, the two policy distributions can be expressed as $\pi^1(a^1_t|s_t)$ and $\pi^2(a^2_t|s_t)$.   Then, for given environment state $s_t$ observed by both agents, $\pi^1(a^1_t|s_t)$ and $\pi^2(a^2_t|s_t)$ are conditionally independent and  the mutual information  $I(\pi^1(\cdot|s_t);\pi^2(\cdot|s_t)) = 0$.
Thus, the MMI objective (\ref{eq:obj_mmi}) reduces to the standard MARL objective of only the accumulated return.
In the following subsections, we present our approach to circumvent this difficulty and implement the MMI framework and its operation under CTDE.

\subsection{Inducing Mutual Information Using Latent Variable}\label{subsec:latentvariable}

First, in order to induce mutual information among agents' policies under the considered system causal diagram shown in Fig. \ref{fig:twoagents}(a),
we introduce  latent variable $z_t$. For illustration, consider the new diagram with latent variable $z_t$
 in Fig. \ref{fig:twoagents}(b). Suppose that the latent variable $z_t$ has a prior distribution $p(z_t)$, and assume that both actions $a_t^1$ and $a_t^2$ are generated from the {\em observed} random variable $s_t$ and the {\em unobserved} random variable $z_t$.
 Then, the policy of agent $i$ is given by the marginal distribution $\pi^i(\cdot|s_t)=\int_z \pi^i(\cdot|s_t,z)p(z)dz$ marginalized over $z$. With the unobserved latent random variable $z$, the conditional independence does not hold for $a_t^1$ and $a_t^2$ and the mutual information can be positive, i.e.,  $I(\pi^1(\cdot|s_t);\pi^2(\cdot|s_t)) > 0$. Hence, we can induce the mutual information between actions without explicit dependence by introducing the latent variable.
 In the general case of $N$ agents, we have  $\boldsymbol{\pi}(a^1,\cdots,a^N|s)=E_z[\pi^1(a^1|s,z)\cdots \pi^N(a^N|s,z)]$.
Note that in this case  we inject a common latent variable $z$ into all agents' policies.

\subsection{Variational Bound of Mutual Information}
\label{subsec:BasicSetup}

Even with non-trivial mutual information $I(\pi^i(\cdot|s_t);\pi^j(\cdot|s_t))$, it is difficult to directly compute the mutual information.
Note that we need the conditional distribution of $a_t^j$ given $(a_t^i,s_t)$ to compute the mutual information as seen in \eqref{eq:MILB1}, but it is difficult to know the conditional distribution directly.
To circumvent this difficulty, we use a variational distribution $q(a_t^j|a_t^i,s_t)$ to approximate $p(a_t^j|a_t^i,s_t)$ and derive a lower bound on the mutual information $ I(\pi^i(\cdot|s_t);\pi^j(\cdot|s_t))=:I_{ij}(s_t)$ as
\begin{align}
        I_{ij}(s_t)
    &= E_{p(a_t^i,a_t^j|s_t)}\left[ \log \frac{q(a_t^j|a_t^i,s_t)}{p(a_t^j)}\right] + E_{p(a_t^i|s_t)}\left[KL(p(a_t^j|a_t^i,s_t)\|q(a_t^j|a^i,s_t) \right] \nonumber \\
    &\geq H(\pi^j(\cdot|s_t))+ E_{p(a_t^i,a_t^j|s_t)}\left[\log q(a_t^j|a_t^i,s_t) \right],  \label{eq:MILB1}
\end{align}
where the inequality holds because KL divergence is always non-negative. The lower bound becomes tight when $q(a_t^j|a_t^i,s_t)$  approximates $p(a_t^j|a_t^i,s_t)$ well.
Using the symmetry of mutual information, we can rewrite the lower bound as
\begin{align}\label{eq:lb_mi2}
    &I_{ij}(s_t)
    \geq \frac{1}{2} \Big[ H(\pi^i(\cdot|s_t))+ H(\pi^j(\cdot|s_t)) + E_{p(a^i,a^j|s_t)}\left[\log q(a^i|a^j,s_t) + \log q(a^j|a^i,s_t) \right] \Big].
\end{align}
Then, we can maximize the lower bound of mutual information by using the tractable approximation $q(a_t^i|a_t^j,s_t)$.

\subsection{Modified Policy Iteration}
\label{subsec:BasicSetup}

In this subsection, we develop policy iteration for the MMI framework. First,  we replace the original MMI objective function (\ref{eq:obj_mmi}) with the following tractable objective function
based on the variational lower bound (\ref{eq:lb_mi2}):
\begin{align}\label{eq:obj_mmi_bound}
        \hat{J}(\boldsymbol{\pi}, q)
        =E_{\boldsymbol{\pi}}\Bigg[\sum_{t=0}^{\infty} & \gamma^t \Big(  r_t(s_t,\boldsymbol{a_t}) +  \alpha N \sum_{i=1}^{N}H(\pi^i(\cdot|s_t))  + \alpha \sum_{i=1}^{N}\sum_{j\neq i}\log q(a_t^j|a_t^i,s_t) \Big) \Bigg],
\end{align}
where  $q(a_t^j|a_t^i,s_t)$ is the variational distribution to approximate the conditional distribution $p(a_t^j|a_t^i,s_t)$.
Then, we determine the individual objective function $\hat{J}^i(\pi^i,q)$  for agent $i$ as the sum of the terms in \eqref{eq:obj_mmi_bound} associated with agent $i$'s policy $\pi^i$ or action $a_t^i$, given by  $\hat{J}^i(\pi^i,q) =$
\begin{align}\label{eq:obj_mmi_bound_agent}
      E_{\boldsymbol{\pi}}\Bigg[\sum_{t=0}^{\infty}\gamma^t \Big( \underbrace{ r_t(s_t,\boldsymbol{a_t}) + \beta  \cdot H(\pi^i(\cdot|s_t))}_{(a)}  + \frac{\beta}{N}\sum_{j\neq i}\Big[\underbrace{\log q(a_t^i|a_t^j,s_t)+\log q(a_t^j|a_t^i,s_t)}_{(b)}\Big] \Big)\Bigg],
\end{align}
where $\beta = \alpha N$ is the temperature parameter.
Note that maximizing the term (a) in (\ref{eq:obj_mmi_bound_agent}) implies that each agent maximizes the weighted sum of the policy entropy and the return, which can be interpreted as an extension of maximum entropy RL to multi-agent setting.
On the other hand, maximizing the term (b) with respect to $\pi^i$ means that
we update the policy $\pi^i$  so that agent $j$ well predicts agent $i$'s action by the first term in (b) and agent $i$ well predicts agent $j$'s action by the second term in (b).
Thus, the objective function (\ref{eq:obj_mmi_bound_agent}) can be interpreted  as the maximum entropy MARL objective combined with
predictability enhancement for other agents' actions.  Note that predictability is reduced  when actions are uncorrelated. Since the policy entropy term $H(\pi^i(\cdot|s_i))$ enhances individual exploration due to maximum entropy principle \cite{2018Haarnoja} and the term (b) in  (\ref{eq:obj_mmi_bound_agent}) enhances predictability or correlation among agents' actions,  the proposed objective function \eqref{eq:obj_mmi_bound_agent} can be considered as one implementation of the concept of correlated exploration in MARL \cite{2019Mahajan2019}.

Now, in order to learn policy $\pi^i$ to maximize the objective function (\ref{eq:obj_mmi_bound_agent}), we modify the policy iteration in standard RL. For this, we redefine the state and state-action value functions for each agent as follows:
\begin{align}
    &V_i^{\boldsymbol{\pi}}(s)
    \triangleq E_{\boldsymbol{\pi}}\Bigg[ \sum_{t=0}^{\infty}\gamma^t \Big(r_{t} +\beta H(\pi^i(\cdot | s_{t})) +\frac{\beta}{N}\sum_{j\neq i}\log q^{(i,j)}(a_{t}^i,a_{t}^j,s_{t})\Big) \Bigg| s_0=s \Bigg] \label{eq:value_function1} \\
    &Q_i^{\boldsymbol{\pi}}(s,a) \triangleq E_{\boldsymbol{\pi}}\Bigg[ r_0 + \gamma V_i^{\boldsymbol{\pi}}(s_{1}) \Bigg| s_0=s, a_0=a \Bigg], \label{eq:value_function2}
\end{align}
where $q^{(i,j)}(a_t^i,a_t^j,s_t)\triangleq q(a_t^i|a_t^j,s_t)q(a_t^j|a_t^i,s_t)$.
Then, the Bellman operator corresponding to  $V_i^{\boldsymbol{\pi}}$ and $Q_i^{\boldsymbol{\pi}}$ is given by
\begin{align}\label{eq:bellman}
    \mathcal{T}^{\boldsymbol{\pi}}Q_i(s,\boldsymbol{a}) &\triangleq r(s,\boldsymbol{a}) + \gamma E_{s'\sim p}[V_i(s')],
\end{align}
where
\begin{align}\label{eq:bellman2}
    &V_i(s)= E_{\boldsymbol{a}\sim \boldsymbol{\pi}}\Bigg[Q_i(s,\boldsymbol{a})
    -\beta \log \pi^i(a^i|s) + \frac{\beta}{N} \sum_{j\neq i}\log q^{(i,j)}(a^i,a^j,s) \Bigg]
\end{align}

In the policy evaluation step, we compute the value functions defined in (\ref{eq:value_function1}) and (\ref{eq:value_function2}) by applying the modified Bellman operator $\mathcal{T}^{\boldsymbol{\pi}}$ repeatedly to any initial function $Q_i^0$.

\begin{lemma}
(Variational Policy Evaluation). For fixed $\boldsymbol{\pi}$ and the variational distribution $q$, consider the modified Bellman operator $\mathcal{T}^{\boldsymbol{\pi}}$ in (\ref{eq:bellman}) and an arbitrary initial function $Q_i^0:\mathcal{S}\times\mathcal{A}\rightarrow \mathbb{R}$, and define $Q_i^{k+1}=\mathcal{T}^{\boldsymbol{\pi}}Q_i^k$. Then, $Q_i^k$ converges to $Q_i^{\boldsymbol{\pi}}$ defined in (\ref{eq:value_function2}).
\end{lemma}

{\it{Proof}}. See Appendix B.
\vspace{0.3em}

In the policy improvement step, we update the policy and the variational distribution by using the value function evaluated in the policy evaluation step. Here,  each agent updates its policy and variational distribution while keeping other agents' policies fixed as follows: $(\pi^i_{k+1}, q_{k+1}) =$
\begin{align}\label{eq:policy_improvement}
    &  \mathop{\arg\max}_{\pi^i, q} E_{(a^i,a^{-i})\sim (\pi^i, \pi_{k}^{-i})} \Bigg[Q_i^{\boldsymbol{\pi}_k}(s,\boldsymbol{a})
    -\beta \log \pi^i(a^i|s) + \frac{\beta}{N} \sum_{j\neq i}\log q^{(i,j)}(a^i,a^j,s) ) \Bigg],
\end{align}
where $a^{-i} \triangleq \{a^1,\cdots,a^N\} \backslash \{a^i\}$. Then, we have the following lemma regarding the improvement step.

\begin{lemma}
(Variational Policy Improvement). Let $\pi_{new}^i$ and $q_{new}$ be the updated policy and the variational distribution from (\ref{eq:policy_improvement}). Then, $Q_i^{\pi^i_{new}, \pi^{-i}_{old}}(s,\boldsymbol{a})\geq Q_i^{\pi^i_{old}, \pi^{-i}_{old}}(s,\boldsymbol{a})$ for all $(s,\boldsymbol{a}) \in (\mathcal{S}\times \boldsymbol{\mathcal{A}})$.
\end{lemma}
{\it{Proof}}. See Appendix B.
\vspace{0.3em}

The modified policy iteration is defined as  applying the variational policy evaluation and variational improvement steps in an alternating manner.
Each agent trains its policy, critics and the variational distribution to maximize its  objective function (\ref{eq:obj_mmi_bound_agent}).

%In order to apply the modified policy iteration to the continuous domain, we need to consider the function approximators.

\begin{figure*}[t]
\begin{center}
\begin{tabular}{c}
     % uncomment the next lines, and give the right ps files
     \includegraphics[width=0.8\textwidth]{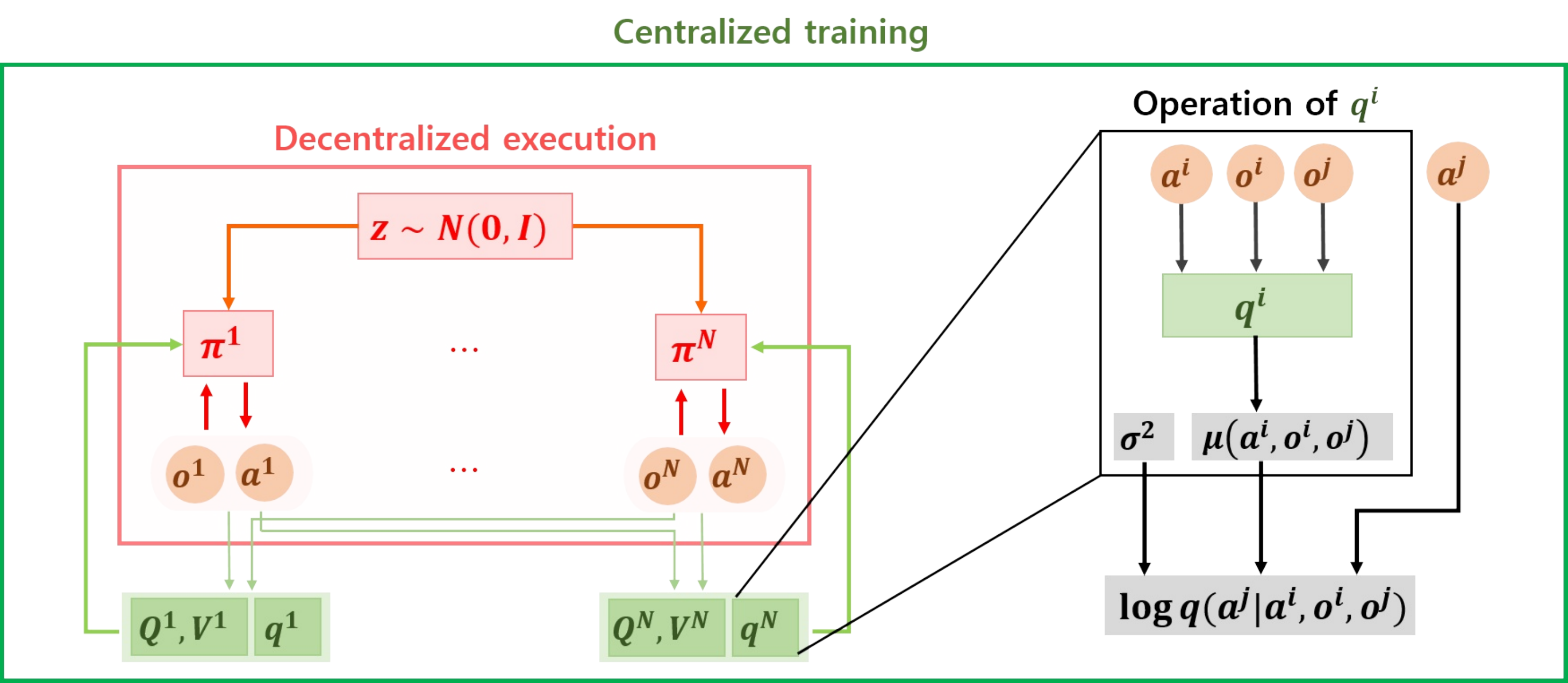}
\end{tabular}
\caption{Overall operation of the proposed VM3-AC. We only need the operation in the red box after training. }
\label{fig:overview}
\end{center}
\end{figure*}

%\begin{wrapfigure}{r}{10cm}
%    \begin{tabular}{@{}c@{}}
%    \includegraphics[width=0.75\textwidth]{figures/fig_mmi7.pdf}
%    \end{tabular}
%    \caption{Overall operation of the proposed VM3-AC. We only need %the operation in the red box after training.}
%    \label{fig:overview}
%\end{wrapfigure}

\section{Algorithm Construction}
\label{sec:BasicSetup}

Summarizing the development above, we now propose the variational maximum mutual information multi-agent actor-critic (VM3-AC) algorithm, which can be applied to  continuous and partially observable multi-agent environments under CTDE.
The overall operation of VM3-AC is shown in Fig. \ref{fig:overview}.
Under CTDE, each agent's policy is conditioned only on local observation, and centralized critics are conditioned on either the environment state or the observations of all agents, depending on the situation \cite{2017Lowe}.
Let $\boldsymbol{x}$ denote either
the environment state $s$ or the observations  of all agents $(o_1,\cdots,o_N)$, whichever is used.
In order to  deal with  the large continuous state-action spaces, we adopt deep neural networks to approximate the required functions.
For agent $i$, we parameterize the variational distribution with $\xi^i$ as $q_{\xi^i}(a^j|a^i,o^i,o^j)$, the state-value function with $\psi^i$ as $V^i_{\psi_i}(\boldsymbol{x})$, two action-value functions with $\theta^{i,1}$ and $\theta^{i,2}$ as $Q^i_{\theta^{i,1}}(\boldsymbol{x},\boldsymbol{a}),Q^i_{\theta^{i,2}}(\boldsymbol{x},\boldsymbol{a})$, and the policy with $\phi^i$ as $\pi_{\phi^i}^i(a|o^i)=E_z[\pi_{\phi^i}^i(a|o^i,z)]$.
We assume normal distribution for the latent variable which plays a key role in inducing coordination among agents' policies,  i.e., $z_t\sim \mathcal{N}(0,I)$, and further assume that the variational distribution is Gaussian distribution with constant variance $\sigma^2$, i.e., $q_{\xi^i}(a^j|a^i,o^i, o^j)=\mathcal{N}(\mu_{\xi^i}(a^i,o^i, o^j), \sigma^2)$, where $\mu_{\xi^i}(a^i,o^i, o^j)$ is the mean of the distribution.

\subsection{Centralized Training}

As aforementioned, the  policy is the marginalized distribution over the latent variable $z$, where the policies of all agents take the same $z_t$ generated from $\mathcal{N}(0,I)$ as an input variable.
We perform the required marginalization based on Monte Carlo numerical expectation as follows:
\begin{align}\label{eq:monte_policy}
    \boldsymbol{\pi}(\boldsymbol{a}|s)
    &= E_z[\pi_{\phi^1}^1(a^1|s,z)\cdots \pi_{\phi^N}^N(a^N|s,z)] \simeq \frac{1}{L}\sum_{l=1}^{L}\pi_{\phi^1}^1(a^1|s,z^l)\cdots \pi_{\phi^N}^N(a^N|s,z^l),
\end{align}
and we use $L=1$ for simplicity.
The value functions $V^i_{\psi_i}(\boldsymbol{x})$, $Q^i_{\theta_i}(\boldsymbol{x},\boldsymbol{a})$ are updated based on the modified Bellman operator defined in (\ref{eq:bellman}) and (\ref{eq:bellman2}). The state-value function $V^i_{\psi_i}(\boldsymbol{x})$ is trained to minimize the following loss function:
\begin{equation}\label{eq:practical_value}
    \mathcal{L}_V(\psi^i)=E_{s_t\sim D}\left[ \frac{1}{2}(V^i_{\psi^i}(\boldsymbol{x}_t)-\hat{V}^i_{\psi^i}(\boldsymbol{x}_t))^2 \right]
\end{equation}
where
$\hat{V}^i_{\psi^i}(\boldsymbol{x}_t) = E_{z\sim N(0,I),\{a^i\sim \pi^i(\cdot|o^i_t,z)\}_{i=1}^{N}}\Bigg[ Q^i_{min}(\boldsymbol{x}_t,\boldsymbol{a}_t)-\beta \log \pi^i_{\phi^i}(a_t^i|o^i_t) + \frac{\beta}{N} \sum_{j\neq i}\log q_{\xi^i}^{(i,j)}(a_t^i,a_t^j,o_t^i,o_t^j) \Bigg]$,
$D$ is the replay buffer that stores the transitions $(\boldsymbol{x}_t,\boldsymbol{a}_t, r_t, \boldsymbol{x}_{t+1})$, and $Q^i_{min}(\boldsymbol{x}_t,a_t^i) = \text{min}[Q^i_{\theta^{i,1}}(\boldsymbol{x}_t,a_t^i), Q^i_{\theta^{i,2}}(\boldsymbol{x}_t,a_t^i)]$ is the minimum of the two action-value functions to prevent the overestimation problem \cite{2018Fujimoto}. The two action-value functions are updated by minimizing the loss
\begin{equation}\label{eq:loss_critic1}
    \mathcal{L}_Q(\theta^i)=E_{(\boldsymbol{x}_t,\boldsymbol{a}_t)\sim D}\left[ \frac{1}{2}(Q_{\theta^i}(\boldsymbol{x}_t,\boldsymbol{a}_t)-\hat{Q}(\boldsymbol{x}_t,\boldsymbol{a}_t))^2 \right]
\end{equation}
where
\begin{equation}\label{eq:loss_critic2}
    \hat{Q}(\boldsymbol{x}_t,\boldsymbol{a}_t) =r_t(x_t,\boldsymbol{a_t}) + \gamma E_{\boldsymbol{x}_{t+1}}[V_{\overline{\psi}^i}\boldsymbol({x}_{t+1})]
\end{equation}
and $V_{\overline{\psi}^i}$ is the target value network, which is updated by the exponential moving average method. We implement the reparameterization trick to estimate the stochastic gradient of policy loss. Then, the action of agent $i$ is given by $a^i=f_{\phi^i}(s;\epsilon^i,z)$, where $\epsilon^i \sim \mathcal{N}(0,I)$ and $z\sim \mathcal{N}(0,I)$. The policy for agent $i$ and the variational distribution are trained to minimize the following policy improvement loss,
\begin{align}\label{eq:loss_actor_va}
        &\mathcal{L}_{\pi^i, q}(\phi^i, \xi)=E_{{\scriptsize \begin{array}{c}
        s_t\sim D,\\
        \epsilon^i\sim \mathcal{N},\\ z\sim \mathcal{N}
        \end{array}}}\Bigg[ -Q^i_{\theta^{i,1}}(\boldsymbol{x}_t,\boldsymbol{a}) + \beta \log \pi^i_{\phi^i}({a}^i|o^i_t) - \frac{\beta}{N} \sum_{j\neq i}\log q_{\xi^i}^{(i,j)}({a}^i,{a}^j,o_t^i,o_t^j) \Bigg]
\end{align}
where
\begin{equation} \label{eq:qxiiij}
q_{\xi^i}^{(i,j)}(a_t^i,a_t^j,o_t^i,o_t^j) = \underbrace{q_{\xi^i}(a_t^i|a_t^j,o_t^i,o_t^j)}_{(a)} \underbrace{q_{\xi^i}(a_t^j|a_t^i,o_t^i,o_t^j)}_{(b)}.
\end{equation}
%Note that from the perspective of agent $i$, maximizing the term (a) in  \eqref{eq:qxiiij} tries to make agent $j$ well predict agent $i$'s action whereas maximizing the term (b) in  \eqref{eq:qxiiij} tries to make agent $i$ itself well predict agent $j$'s action.
Since approximation of the variational distribution is not accurate in the early stage of training and the learning via the term (a) in  \eqref{eq:qxiiij} is more susceptible to approximation error, we propagate the gradient only through the term (b)  in  \eqref{eq:qxiiij} to  make learning stable.
Note that minimizing $-\log q_{\xi^i}(a^j|a^i,s_t)$ is equivalent to minimizing the mean-squared error between $a^j$ and $\mu_{\xi^i}(a^i,o^i,o^j)$ due to our Gaussian assumption on the variational distribution.

\subsection{Decentralized Execution}\label{subsec:execution}

In the centralized training phase, we pick the actions $(a^1,\cdots,a^N)$  by using  Monte Carlo expectation  based on common  latent variable $z^l$ generated from zero-mean Gaussian distribution, as seen in \eqref{eq:monte_policy}. We can also achieve the same operation in the decentralized execution phase.
This can be done  by making all agents have the same Gaussian random sequence generator and distributing the same seed to this random sequence generator only once in the beginning of the execution phase. This  eliminates the necessity of communication for sharing the latent variable. In fact, this
way of sharing $z^l$ can be applied to the centralized training phase too.
The proposed VM3-AC algorithm is summarized in Appendix C.

%%%%%%%%%%%%%%%%%%%%%%%%%%%%%%%%%%%%%%%%%%%%%%%%%%%%%%%%%%%%%%%%%%%%%%%%%%%%%%%%%%%
%%%%%%%%%%%%%%%%%%%%%%%%%%%%%%%%%%%%%%%%%%%%%%%%%%%%%%%%%%%%%%%%%%%%%%%%%%%%%%%%%%%
\section{Experiment}
\label{sec:Experiment}
%%%%%%%%%%%%%%%%%%%%%%%%%%%%%%%%%%%%%%%%%%%%%%%%%%%%%%%%%%%%%%%%%%%%%%%%%%%%%%%%%%%

In this section, we provide numerical results to evaluate VM3-AC. Since we focus on the continuous action-space case in this paper, we considered four baselines relevant to the continuous action-space case: 1) MADDPG \cite{2017Lowe} - an extension of DDPG with a centralized critic to train a decentralized policy for each agent. 2) I-SAC - an example of independent learning where each agent learns policy based on SAC while treating other agents as a part of the environment. 3) MA-SAC -  an extension of I-SAC with a centralized critic instead of a decentralized critic. 4) Multi-agent actor-critic (MA-AC) - a variant of MA-SAC, i.e., the same algorithm with MA-SAC without the entropy term.
 All algorithms used  neural networks to approximate the required functions.
 In the algorithms except I-SAC, we used the neural network architecture proposed in \cite{2019Kim} to emphasize the agent's own observation and action for centralized critics. For agent $i$, we used the shared neural network for the variational distribution $q_{\xi^i}(a_t^j|a_t^i,o_t^i, o_t^j )$ for $j\in \{1,\cdots,N\} \backslash \{i\}$, and the network takes the one-hot vector which indicates $j$ as input. Experimental details are given in Appendix E.

We evaluated the proposed algorithm and the baselines in the three multi-agent environments with varying number of agents: multi-walker \cite{2018Gupta}, predator-prey \cite{2017Lowe}, and cooperative navigation \cite{2017Lowe}. The detailed setting of each environments is provided in Appendix D.

%\tcr{The three environments for evaluating our algorithm are multi-walker \cite{2018Gupta}, predator-prey and cooperative navigation \cite{2017Lowe}. The settings of the experiments are slightly different from the those in \cite{2018Gupta} and \cite{2017Lowe} and we provide the detail settings in Appendix.}

\begin{figure*}[t]
\begin{center}
\begin{tabular}{ccc}
     % uncomment the next lines, and give the right ps files
     %\includegraphics[width=0.23\textwidth]{figures/MW_N2.png} &
     \includegraphics[width=0.28\textwidth]{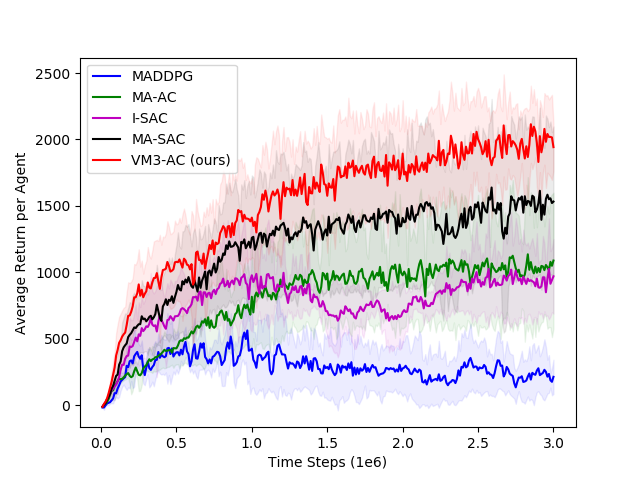} &
     \includegraphics[width=0.28\textwidth]{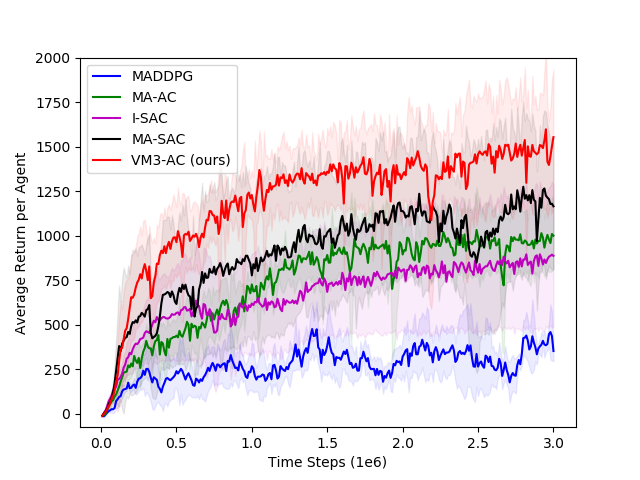} &
     \includegraphics[width=0.28\textwidth]{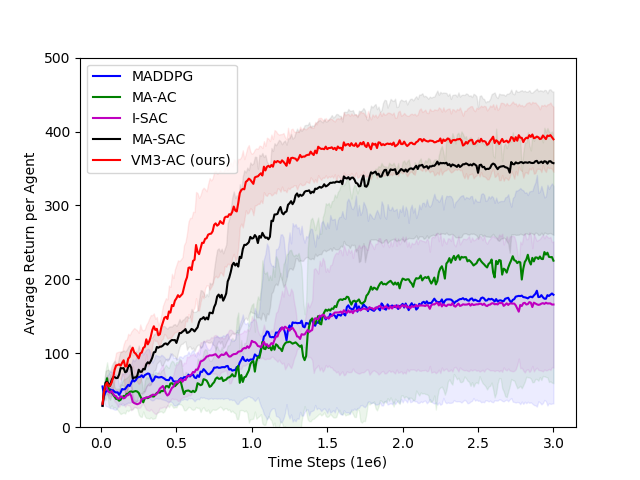} \\
     (a) MW (N=3) & (b) MW (N=4) & (c) PP (N=2)\\
     \includegraphics[width=0.28\textwidth]{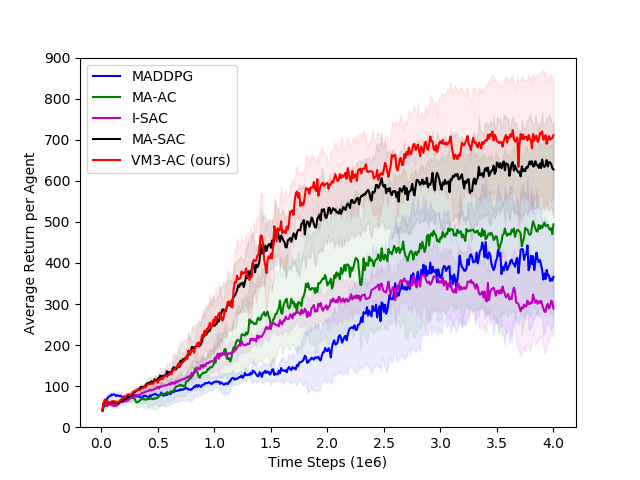} &
     \includegraphics[width=0.28\textwidth]{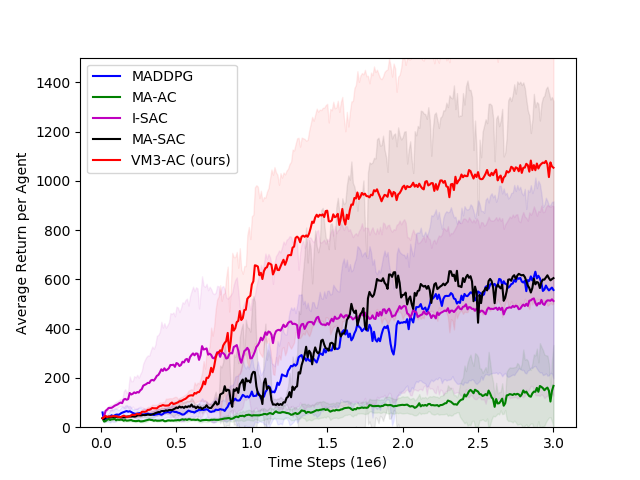} &
     \includegraphics[width=0.28\textwidth]{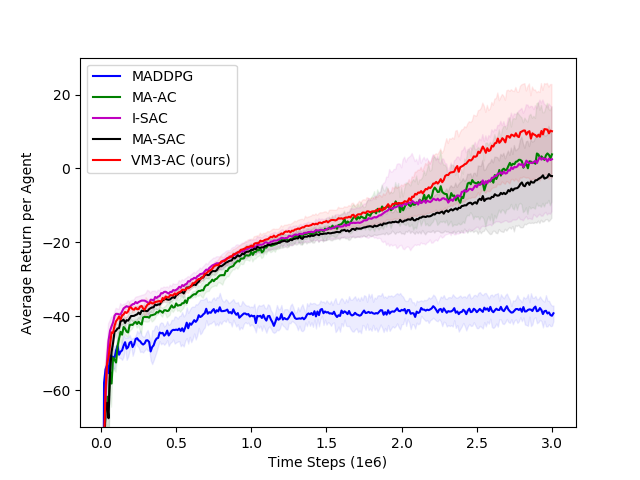}\\
     (d) PP (N=3) & (e) PP (N=4) & (f) CN (N=3)
\end{tabular}
\caption{Performance for MADDPG (blue), MA-AC (green), I-SAC (purple), MA-SAC (black), and VM3-AC (the proposed method, red) on multi-walker environments (a)-(b), predator-prey (c)-(e), and cooperative navigation (f). (MW, PP, and CN denote multi-walker, predator-prey, and cooperative navigation environments, respectively)}
\label{fig:results}
\end{center}
\end{figure*}

%\begin{figure*}[t]
%\begin{center}
%\begin{tabular}{cccc}
     % uncomment the next lines, and give the right ps files
%     \includegraphics[width=0.23\textwidth]{figures/MW_N2.png} &
%     \includegraphics[width=0.23\textwidth]{figures/MW_N3.png} &
%     \includegraphics[width=0.23\textwidth]{figures/MW_N4.png} &
%     \includegraphics[width=0.23\textwidth]{figures/PP_N2.png} \\
%     (a) MW (N=2) & (b) MW (N=3) & (c) MW (N=4) & (d) PP (N=2)\\
%     \includegraphics[width=0.23\textwidth]{figures/PP_N3.png} &
%     \includegraphics[width=0.23\textwidth]{figures/PP_N4.png} &
%     \includegraphics[width=0.23\textwidth]{figures/Cooperative_nagivation_n3.png} &
%     \includegraphics[width=0.23\textwidth]{figures/Cooperative_nagivation_n4.png} \\
%     (e) PP (N=3) & (f) PP (N=4) & (g) CN (N=3) & (h) CN (N=4)
%\end{tabular}
%\caption{Performance for MADDPG (blue), MA-AC (green), I-SAC (purple), MA-SAC (black), and VM3-AC (the proposed method, red) on multi-walker environments (a)-(c), predator-prey (d)-(f), and cooperative navigation (g)-(h). (MW, PP, and CN denote multi-walker, predator-prey, and cooperative navigation environments, respectively)}
%\label{fig:results}
%\end{center}
%\end{figure*}

\subsection{Result}

Fig. \ref{fig:results} shows the learning curves for the considered three environments with the different number of agents. The y-axis denotes the average of all agents' rewards averaged over 7 random seeds, and the x-axis denotes time step. The hyperparameters including the temperature parameter $\beta$ and the dimension of the latent variable are provided in Appendix E.

As shown in Fig. \ref{fig:results}, VM3-AC outperforms the baselines in the considered environments. Especially, in the case of the multi-walker environment, the proposed VM3-AC algorithm has large performance gain. This is because the agents in the multi-walker environment are required especially to learn coordinated behavior to obtain high rewards.
Hence, we can see that  the proposed MMI framework improves  performance in complex multi-agent tasks requiring high-quality coordination. The performance gap between VM3-AC and MA-SAC indicates
the effect of regularization with  the variational term (b) of the objective function (\ref{eq:obj_mmi_bound_agent}). Recall that VM3-AC without the variational term (b) of the objective function (\ref{eq:obj_mmi_bound_agent}) reduces to MA-SAC. Recall also that MA-SAC without entropy regularization reduces to MA-AC, and MA-SAC with decentralized critics instead of centralized critics reduces to I-SAC. Hence, regularization with  entropy and use of centralized critics are also important in  multi-agent tasks from the fact that MA-SAC outperforms I-SAC and MA-AC.  Note that VM3-AC also maximizes the entropy through the term (a) of the objective function (\ref{eq:obj_mmi_bound_agent}).
Indeed, it is seen that regularization with the variation term in addition to policy entropy enhances coordinated behavior in MARL.

\begin{figure*}[h]
\begin{center}
\begin{tabular}{cccc}
     % uncomment the next lines, and give the right ps files
     \includegraphics[width=0.23\textwidth]{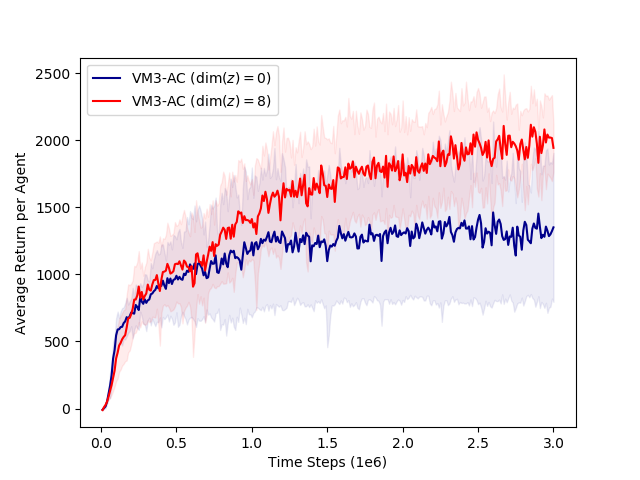} &
     \includegraphics[width=0.23\textwidth]{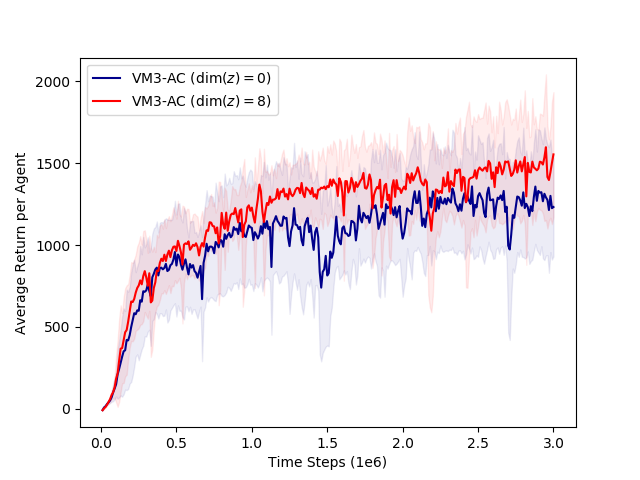} &
     \includegraphics[width=0.23\textwidth]{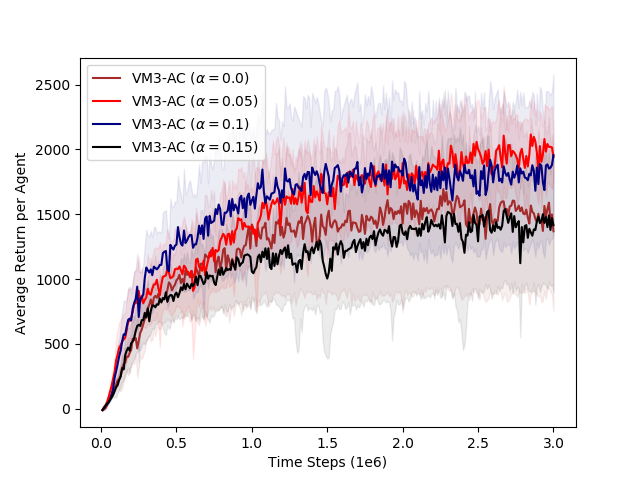} &
     \includegraphics[width=0.23\textwidth]{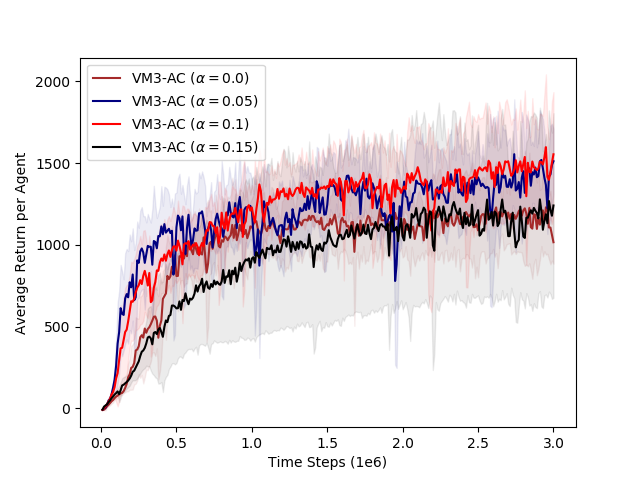} \\
     (a) MW (N=3) & (b) MW (N=4) & (c) MW (N=3) & (d) MW (N=4)
\end{tabular}
\caption{(a) \& (b): Impact of the latent variable  and (c) \& (d): impact of the temperature parameter $\beta$}
\label{fig:ablations}
\end{center}
\end{figure*}

\vspace{-1em}
Due to the space limitation, more result on comparison with the latest algorithm MAVEN \cite{2019Mahajan2019} is provided in Appendix F. It is seen there that VM3-AC significantly outperforms MAVEN.

\subsection{Ablation Study}

In this section, we provide ablation study on the major techniques and hyperparameter of VM3-AC: 1) the latent variable, and 2) the temperature parameter $\beta$.

\textbf{Latent variable:} The role of the latent variable is to induce mutual information among actions and promote coordinated behavior. We compared VM3-AC and VM3-AC without the latent variable (implemented by setting $\mbox{dim}(z)=0$)  in the multi-walker environment with $N=3$ and $N=4$. In both cases, VM3-AC yields better performance that VM3-AC without the latent variable as shown in Fig.\ref{fig:ablations}(a) and  \ref{fig:ablations}(b).

\textbf{Temperature parameter $\beta$:} The role of temperature parameter $\beta$ is to control the relative importance between the reward and the mutual information. We evaluated VM3-AC by varying  $\beta=[0, 0.05, 0.1, 0.15]$ in the multi-walker environment with $N=3$ and $N=4$. Fig. \ref{fig:ablations}(c) and \ref{fig:ablations}(d) show that VM3-AC with the temperature value around $[0.05, 0.1]$ yields good performance.

\section{Conclusion}

\label{sec:Conclusion}

In this paper, we have proposed the MMI framework for MARL  to enhance multi-agent coordinated learning under CTDE by regularizing the cumulative return with mutual information among actions. The MMI framework is implemented practically by using a latent variable and variational technique and applying policy iteration. Numerical results show that the derived algorithm named VM3-AC outperforms other baselines, especially in multi-agent tasks requiring high coordination among agents. Furthermore, the MMI framework can be combined with the other techniques for cooperative MARL, such as value decomposition \cite{2018rashid} to yield better performance.

\newpage


%\bibliography{MMI}
%\bibliographystyle{icml2020}
%%%%%%%%%%%%%%%%%%%%%%%%%%%%%%%%%

%\bibliographystyle{neurips_2020}
\section*{Broader Impact}

The research topic of this paper is multi-agent reinforcement learning (MARL). MARL is an important branch in the field of reinforcement learning.
MARL models many of practical control problems in the real world such as smart factories, coordinated robots and connected self-driving cars. With the advance of knowledge and technologies in MARL, solutions to such real-world problems can be improved and more robust. For example, if the control of self-driving cars are coordinated among several near-by cars, the safety involved in self-driving cars will be improved much. So, we believe that the research advances in this field can benefit our safety and future society.

%Authors are required to include a statement of the broader impact of their work, including its ethical aspects and future societal consequences.
%Authors should discuss both positive and negative outcomes, if any. For instance, authors should discuss a)

%who may benefit from this research, b) who may be put at disadvantage from this research, c) what are the consequences of failure of the system, and d) whether the task/method leverages
%biases in the data. If authors believe this is not applicable to them, authors can simply state this.
%Use unnumbered first level headings for this section, which should go at the end of the paper. {\bf Note that this section does not count towards the eight pages of content that are allowed.}

\nocite{langley00}

\bibliography{MMI}
\bibliographystyle{icml2020}

\newpage

%%%%%%%%%%%%%%%%%%%%%%%%%%%%%%%%%%%%%%%%%%%%%%%%%%%%%%%%%%%%%%%%%%%%%%%%%%%%%%
\section*{Appendix A: Related Work}
\label{sec:Relatedwork}

For cooperative MARL, several approaches have been studied. One of the approaches is value decomposition techniques \cite{2017sunehag, 2018rashid, 2019son}. For example, QMIX \cite{2018rashid} factorizes the joint action-value function into a combination of local action-value functions while imposing a monotonicity constraint. QMIX achieves  state-of-the-art performance in complex discrete-action MARL tasks and has been widely used as a baseline in discrete-action environments.
 Since the focus of VM3-AC  is on continuous-action environments, the direct comparison of VM3-AC to QMIX is irrelevant. However, the basic concept of QMIX can also be applied to the MMI framework, and this remains  as future work.

Learning coordinated behavior in the multi-agent systems is studied extensively in the MARL community.
To promote coordination, some previous works used communication among agents \cite{2013zhang, 2016foerester, 2019Pesce}. For example, \cite{2016foerester} proposed the DIAL algorithm  to learn  communication protocol that enables the agents to coordinate their behaviors.
\cite{2019Jaques} proposed the social influence intrinsic reward which is related to the mutual information between actions to  achieve coordination. Although the social influence algorithm increases the performance in challenging social dilemma environments, the limitation is that explicit dependency across  actions is required and imposed for this algorithm to compute the intrinsic reward.
As already mentioned, the MMI framework can be viewed as indirect enhancement of correlated exploration.  The correlated policies are considered in several other works too. \cite{2020Liu} proposed the explicit modeling of correlated policies for multi-agent imitation learning, and \cite{2019Wen} proposed a probabilistic recursive reasoning framework.
By introducing a latent variable and variational lower bound on mutual information, the proposed VM3-AC increases the correlation among policies without communication in the execution phase and without explicit dependency across agents' actions.

As mentioned in the main paper, the proposed MMI framework can be interpreted as enhancing correlated exploration by increasing the entropy of own policy while decreasing the uncertainty about other agents' actions. Some previous works also proposed other techniques to enhance  correlated exploration \cite{2019Mahajan2019, 2018zheng}. For example, MAVEN addressed the poor exploration of QMIX by maximizing the mutual information between the latent variable and the observed trajectories \cite{2019Mahajan2019}. However, MAVEN does not consider the correlation among policies. We compare the proposed VM3-AC with MAVEN and the comparison result is given in Appendix F.

%%%%%%%%%%%%%%%%%%%%%%%%%%%%%%%%%%%%%%%%%%%%%%%%%%%%%%%%%%%%%%%%%%%%%%%%%%%%%%
\newpage
\section*{Appendix B: Variational policy evaluation and policy improvement}

In the main paper, we defined the state and state-action value functions for each agent as follows:
\begin{align}
    &V_i^{\boldsymbol{\pi}}(s)
    \triangleq E_{\boldsymbol{\pi}}\Bigg[ \sum_{t=0}^{\infty}\gamma^t \Big(r_{t} +\beta H(\pi^i(\cdot | s_{t}))+\frac{\beta}{N}\sum_{j\neq i}\log q^{(i,j)}(a_{t}^i,a_{t}^j,s_{t})\Big) \Bigg| s_0=s \Bigg] \label{eq:value_function1} \\
    &Q_i^{\boldsymbol{\pi}}(s,a) \triangleq E_{\boldsymbol{\pi}}\Bigg[ r_0 + \gamma V_i^{\boldsymbol{\pi}}(s_{1}) \Bigg| s_0=s, a_0=a \Bigg], \label{eq:value_function2}
\end{align}

\begin{lemma}
(Variational Policy Evaluation). For fixed $\boldsymbol{\pi}$ and the variational distribution $q$, consider the modified Bellman operator $\mathcal{T}^{\boldsymbol{\pi}}$ in (\ref{eq:bellman}) and an arbitrary initial function $Q_i^0:\mathcal{S}\times\mathcal{A}\rightarrow \mathbb{R}$, and define $Q_i^{k+1}=\mathcal{T}^{\boldsymbol{\pi}}Q_i^k$. Then, $Q_i^k$ converges to $Q_i^{\boldsymbol{\pi}}$ defined in (\ref{eq:value_function2}).
\end{lemma}

\begin{align}\label{eq:bellman}
    \mathcal{T}^{\boldsymbol{\pi}}Q_i(s,\boldsymbol{a}) &\triangleq r(s,\boldsymbol{a}) + \gamma E_{s'\sim p}[V_i(s')],
\end{align}
where
\begin{align}\label{eq:bellman2}
    &V_i(s)= E_{\boldsymbol{a}\sim \boldsymbol{\pi}}\Bigg[Q_i(s,\boldsymbol{a})
    -\beta \log \pi^i(a^i|s) + \frac{\beta}{N} \sum_{j\neq i}\log q^{(i,j)}(a^i,a^j,s) \Bigg]
\end{align}

\begin{proof}
	Define the mutual information augmented reward as $\mathcal{T}^{\pi}Q_i(s_t,\boldsymbol{a_t})=$
	\begin{align}
    	&= r(s_t,\boldsymbol{a_t})     +\gamma E_{s_{t+1}\sim p, \boldsymbol{a_{t+1}}\sim \boldsymbol{\pi}} \Bigg[Q_i(s_{t+1},\boldsymbol{a_{t+1}})-\beta \log \pi^i(a_t^i|s_t) + \frac{\beta}{N} \sum_{j\neq i}\log q^{(i,j)}(a_t^i,a_t^j,s_t) \Bigg] \\
	    &= \underbrace{ r(s_t,\boldsymbol{a_t})     +\gamma  E_{s_{t+1}\sim p, \boldsymbol{a_{t+1}}\sim \boldsymbol{\pi}}\Bigg[-\beta \log \pi^i(a_t^i|s_t) + \frac{\beta}{N} \sum_{j\neq i}\log q^{(i,j)}(a_t^i,a_t^j,s_t)\Bigg]}_{r_{\pi}(s_t,\boldsymbol{a_t})} \\
	    &+\gamma E_{s_{t+1}\sim p, \boldsymbol{a_{t+1}}\sim \boldsymbol{\pi}} \Bigg[Q_i(s_{t+1},\boldsymbol{a_{t+1}})\Bigg]  \\
	    &= r_{\pi}(s_t,\boldsymbol{a_t})+\gamma E_{s_{t+1}\sim p, \boldsymbol{a_{t+1}}\sim \boldsymbol{\pi}}\Bigg[Q_i(s_{t+1},\boldsymbol{a_{t+1}})\Bigg]
	\end{align}
	Then, we can apply the standard convergence results for policy evaluation. Define
	\begin{align}
    	\mathcal{T}^{\pi}(v) = \mathcal{R}^{\pi} + \gamma \mathcal{P}^{\pi}v
	\end{align}
	for $v = [Q(s,\boldsymbol{a})]_{s \in \mathcal{S}, \boldsymbol{a} \in \mathcal{A}}$.
	Then, the operator $\mathcal{T}^{\pi}$ is a $\gamma$-contraction.
	\begin{align}
    	\|\mathcal{T}^{\pi}(v) - \mathcal{T}^{\pi}(u)\|_{\infty}
    	&= \| (\mathcal{R}^{\pi} + \gamma \mathcal{P}^{\pi}v) - (\mathcal{R}^{\pi} + \gamma \mathcal{P}^{\pi}u) \|_{\infty} \\
	    &= \| \gamma\mathcal{P}^{\pi}(v-u) \|_{\infty} \\
    	&\leq \|  \gamma\mathcal{P}^{\pi}\|_{\infty}   \| v-u\|_{\infty} \\
    	&\leq \gamma \| u-v \|_{\infty}
	\end{align}
	Note that the operator $\mathcal{T}^{\pi}$ has an unique fixed point by the contraction mapping theorem, and we define the fixed point as $Q_i^\pi(s,\boldsymbol{a})$. Since
	\begin{equation}
		\| Q_i^{k}(s,\boldsymbol{a}) - Q_i^\pi(s,\boldsymbol{a}) \|_{\infty} \leq \gamma \| Q_i^{k-1}(s,\boldsymbol{a}) - Q_i^\pi(s,\boldsymbol{a})  \|_{\infty} \leq \cdots \leq \gamma^{k} \| Q_i^{0}(s,\boldsymbol{a}) - Q_i^\pi(s,\boldsymbol{a}) \|_{\infty},
	\end{equation}
	we have
	\begin{align}
		\lim_{k \rightarrow \infty} \| Q_i^k (s,\boldsymbol{a}) - Q_i^\pi(s,\boldsymbol{a}) \|_\infty = 0
	\end{align}
	and this implies
	\begin{equation}
		\lim_{k \rightarrow \infty} Q_i^k (s,\boldsymbol{a}) = Q_i^\pi(s,\boldsymbol{a}), \quad \forall (s, \boldsymbol{a}) \in (\mathcal{S} \times \boldsymbol{\mathcal{A}}).
	\end{equation}
	
\end{proof}

%\begin{theorem}
%(Contraction Mapping Theorem).
%For any metric space $\mathcal{V}$ that is complete under an operator $T(v)$, where $\mathcal{T}$ is a $\gamma$-contraction,
%\begin{itemize}
%    \item $\mathcal{T}$ converges to a unique fixed point
%    \item At a linear convergence rate of $\gamma$
%\end{itemize}
%\end{theorem}

\vspace{3ex}

\begin{lemma}
(Variational Policy Improvement). Let $\pi_{new}^i$ and $q_{new}$ be the updated policy and the variational distribution from (\ref{eq:policy_improvement}). Then, $Q_i^{\pi^i_{new}, \pi^{-i}_{old}}(s,\boldsymbol{a})\geq Q_i^{\pi^i_{old}, \pi^{-i}_{old}}(s,\boldsymbol{a})$ for all $(s,\boldsymbol{a}) \in (\mathcal{S}\times \boldsymbol{\mathcal{A}})$.
\end{lemma}

\begin{align}\label{eq:policy_improvement}
    & (\pi^i_{k+1}, q_{k+1}) = \mathop{\arg\max}_{\pi^i, q} E_{(a^i,a^{-i})\sim (\pi^i, \pi_{k}^{-i})} \Bigg[Q_i^{\boldsymbol{\pi}_k}(s,\boldsymbol{a}) -\beta \log \pi^i(a^i|s) \\
    &~~~~~~~~~~~~~~~~~~~~~~~~~~~~~~~~~~~~~+ \frac{\beta}{N} \sum_{j\neq i}\log q^{(i,j)}(a^i,a^j,s) ) \Bigg],
\end{align}

\begin{proof}
	Let $\pi_{new}$ be determined as
	\begin{align}
    	(\pi^i_{new}, q_{new}) &= \mathop{\arg\max}_{\pi^i,q} E_{(a_t^i,a_t^{-i})\sim (\pi^i, \pi_{old}^{-i})} \Bigg[Q_i^{\boldsymbol{\pi}_{old}}(s_t,\boldsymbol{a}_t) -\beta \log \pi^i(a_t^i|s_t) \\
    	&~~~~~~~~~~~~~~~+ \frac{\beta}{N} \sum_{j\neq i}\log q^{(i,j)}(a_t^i,a_t^j,s_t) ) \Bigg].
	\end{align}
	Then, the following inequality is hold
	\begin{align}
    	&E_{(a_t^i,a_t^{-i})\sim (\pi_{new}^i, \pi_{old}^{-i})} \Bigg[Q_i^{\boldsymbol{\pi}_{old}}(s_t,\boldsymbol{a}_t) -\beta \log \pi_{new}^i(a_t^i|s_t) + \frac{\beta}{N} \sum_{j\neq i}\log q^{(i,j)}_{new}(a_t^i,a_t^j,s_t) ) \Bigg] \\
    	&\geq E_{(a_t^i,a_t^{-i})\sim (\pi^i_{old}, \pi_{old}^{-i})} \Bigg[Q_i^{\boldsymbol{\pi}_{old}}(s_t,\boldsymbol{a}_t) -\beta \log \pi_{old}^i(a_t^i|s_t) + \frac{\beta}{N} \sum_{j\neq i}\log q^{(i,j)}_{old}(a_t^i,a_t^j,s_t) ) \Bigg] \\
    	&= V^{\boldsymbol{\pi}_{old}}_i(s_t).
	\end{align}

	From the definition of the Bellman operator,
	\begin{align}
    	Q_i^{\boldsymbol{\pi}_{old}}(s_t,\boldsymbol{a_t})
    	&= r(s_t,\boldsymbol{a_t}) + \gamma E_{s_{t+1}\sim p}[V_i^{\boldsymbol{\pi}_{old}}(s_{t+1})] \\
	    &\leq r(s_t,\boldsymbol{a_t}) + \gamma E_{s_{t+1}\sim p}E_{(a_{t+1}^i,a_{t+1}^{-i})\sim (\pi_{new}^i, \pi_{old}^{-i})} \Bigg[Q_i^{\boldsymbol{\pi}_{old}}(s_{t+1},\boldsymbol{a}_{t+1})  \nonumber \\
	    & \ \ \ \ \ -\beta \log \pi_{new}^i(a_{t+1}^i|s_{t+1}) + \beta \sum_{j\neq i}\log q_{new}^{(i,j)}(a_{t+1}^i,a_{t+1}^j,s_{t+1})  \Bigg] \\
	    &\leq r(s_t,\boldsymbol{a_t}) + \gamma E_{s_{t+1}\sim p}E_{(a_{t+1}^i,a_{t+1}^{-i})\sim (\pi_{new}^i, \pi_{old}^{-i})} \Bigg[r^i(s_{t+1},\boldsymbol{a_{t+1}})  \nonumber \\
	    & \ \ \ \ \ -\beta \log \pi_{new}^i(a_{t+1}^i|s_{t+1}) + \beta \sum_{j\neq i}\log q_{new}^{(i,j)}(a_{t+1}^i,a_{t+1}^j,s_{t+1})+ \gamma V_i^{\boldsymbol{\pi}_{old}}(s_{t+2})\Bigg] \\
	    &\ \ \ \vdots \nonumber \\
    	&\leq Q_i^{\pi^i_{new}, \pi^{-i}_{old}}(s_t,a_t).
	\end{align}
\end{proof}

\newpage

\section*{Appendix C: Pseudo Code}

\begin{algorithm}[h]
   \caption{VM3-AC (L=1)}
   \label{alg:VM3-AC}
\begin{algorithmic}
   \STATE \textbf{Centralized training phase}
   \STATE Initialize parameter $\phi^i, \theta^i, \psi^i, \overline{\psi}^i,  \xi^i, ~\forall  i\in \{1,\cdots, N\}$
   \FOR{$episode=1,2,\cdots$}
   \STATE Initialize state $s_0$ and each agent observes $o_0^i$
   \FOR{$t<T$ and $s_t \neq$ terminal}
   \STATE Generate $z_t \sim \mathcal{N}(0,I)$ and select action $a_t^i\sim \pi^i(\cdot|o_t^i,z_t)$ for each agent $i$
   \STATE Execute $\boldsymbol{a_t}$ and each agent $i$ receives $r_t$ and $o_{t+1}^i$
   \STATE Store transitions in $D$
   \ENDFOR
   \FOR{each gradient step}
   \STATE Sample a minibatch from D and generate $z_l \sim \mathcal{N}(0,I)$ for each transition.
   \STATE Update $\theta^i, \psi^i$ by minimizing the loss (\ref{eq:loss_critic1}) and (\ref{eq:loss_critic2})
   \STATE Update $\phi^i, \xi^i$ by minimizing the loss (\ref{eq:loss_actor_va})
   \ENDFOR
   \STATE Update $\overline{\psi}^i$ using the moving average method
   \ENDFOR
   \STATE
   \STATE \textbf{Decentralized execution phase}
   \STATE Initialize state $s_0$ and each agent observes $o_0^i$
   \FOR {each environment step}
   \STATE Select action $a_t^i\sim \pi^i(\cdot|o_t^i,z_t)$ where $z_t=\overrightarrow{0}$
   \STATE Execute $\boldsymbol{a_t}$ and each agent $i$ receives $o_{t+1}^i$
   \ENDFOR
\\
\end{algorithmic}
\end{algorithm}

\newpage

\section*{Appendix D: Environment Detail}
\textbf{Multi-walker} The multi-walker environment, which was introduced in \cite{2018Gupta}, is a modified version of the BipedalWalker environment in OpenAI gym  to multi-agent setting. The environment consists of $N$ bipedal walkers and a large package. The goal of the environment is to move forward together while holding the large package on top of the walkers.
The observation of each agent consists of the joint angular speed, the position of joints and so on.
Each agent has 4-dimensional continuous actions that control the torque of their legs.
Each agent receives  shared reward $R_1$ depending on the distance over which the package has moved and receives  negative local compensation $R_2$ if the agent drops the package or falls to the ground.
 An episode ends when one of the agents falls,  the package is dropped or $T$ time steps elapse.
  To obtain higher rewards, the agents should learn coordinated behavior.
  For example, if one agent only tries to learn to move forward, ignoring  other agents, then  other agents may fall.
  In addition, the different coordinated behavior is required as the number of agents changes.
    We set $T=500$, $R_2=-10$ and $R_1=10d$, where $d$ is the distance over which the package has moved.
    We simulated this environment in three cases by changing the number of agents ($N=2$, $N=3$, and $N=4$).

\begin{wrapfigure}{r}{6.5cm}
    \begin{tabular}{@{}ccc@{}}
    \includegraphics[width=0.18\textwidth]{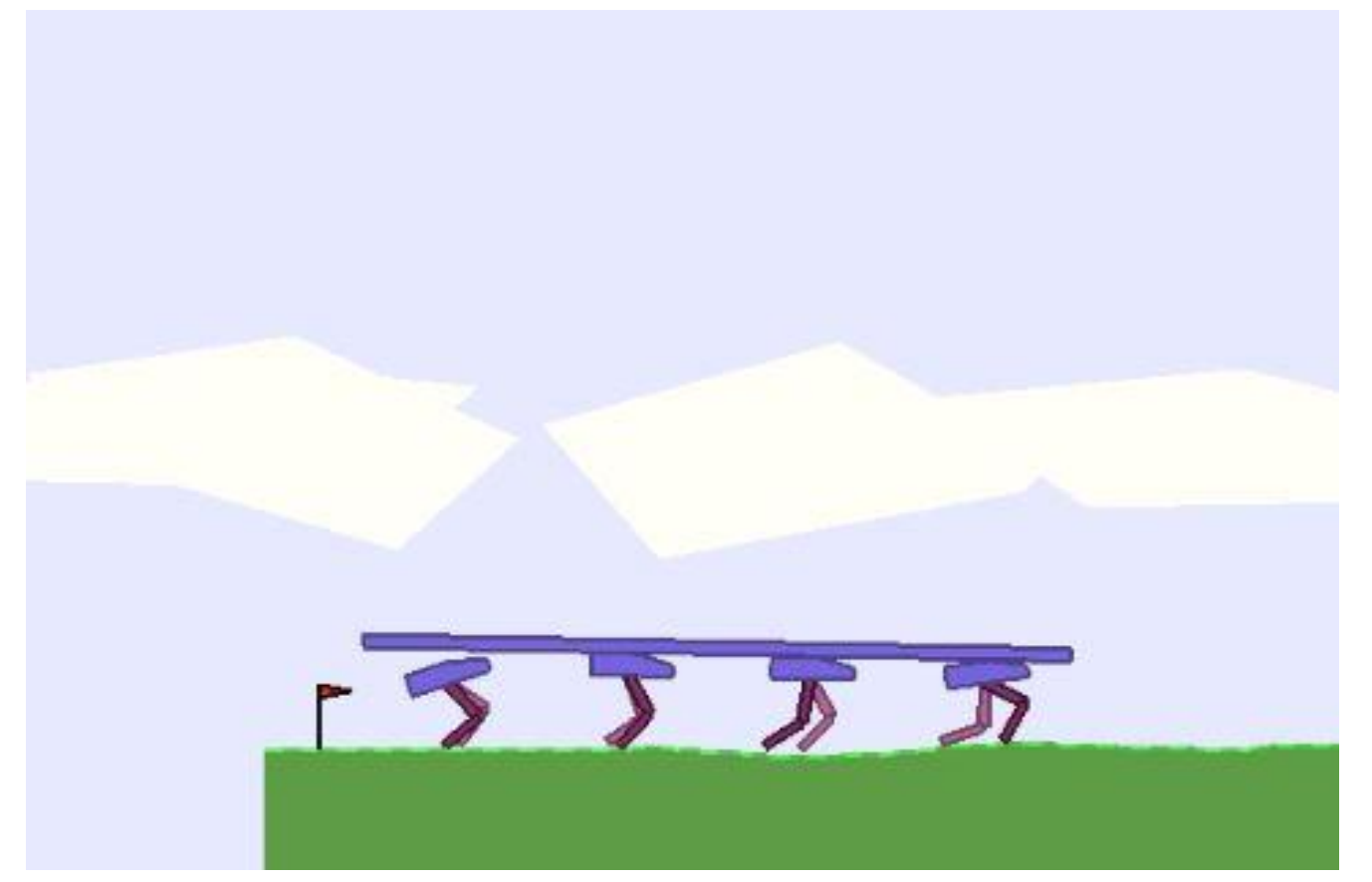} &
     \includegraphics[width=0.13\textwidth]{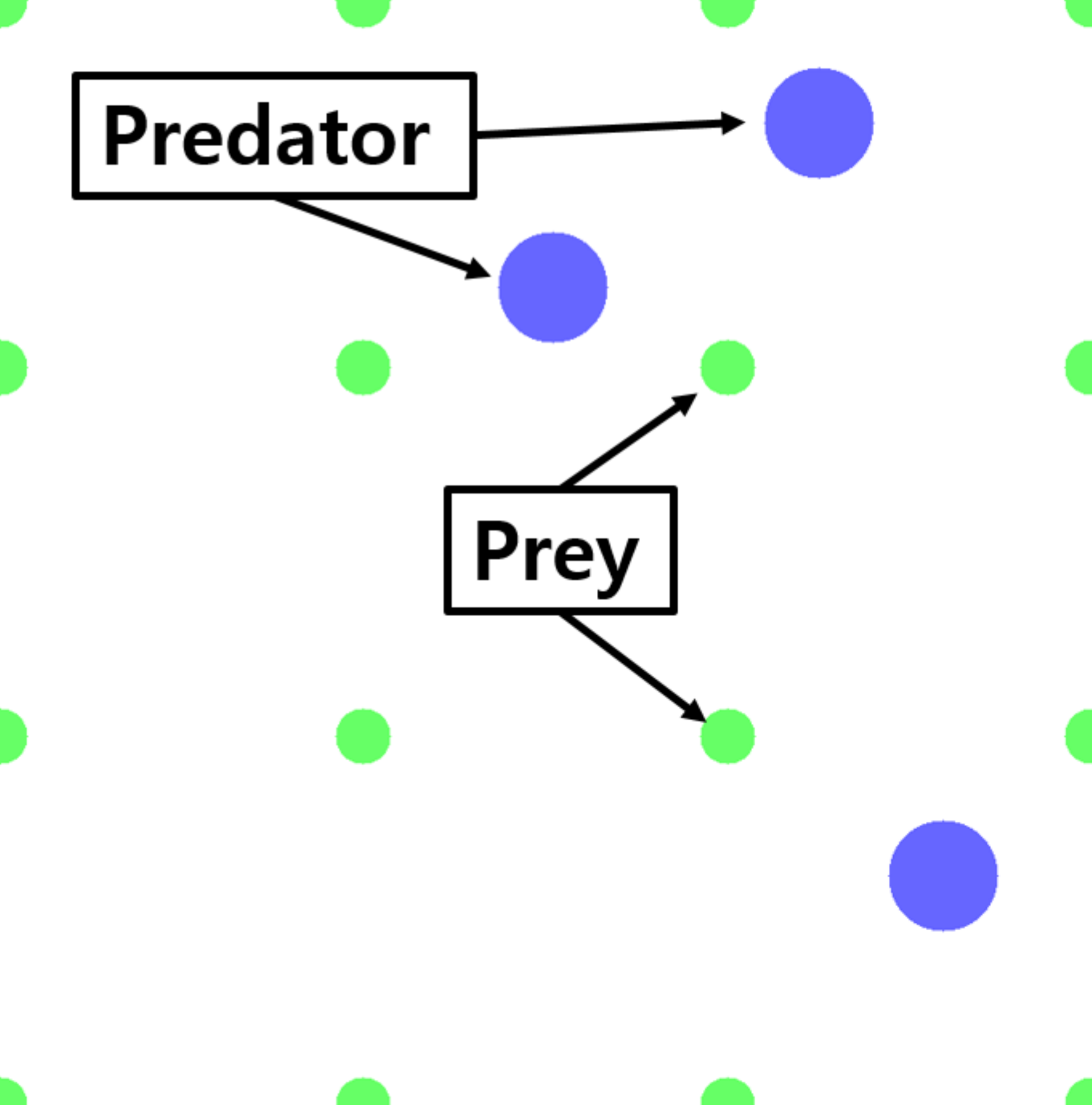} &
     \includegraphics[width=0.12\textwidth]{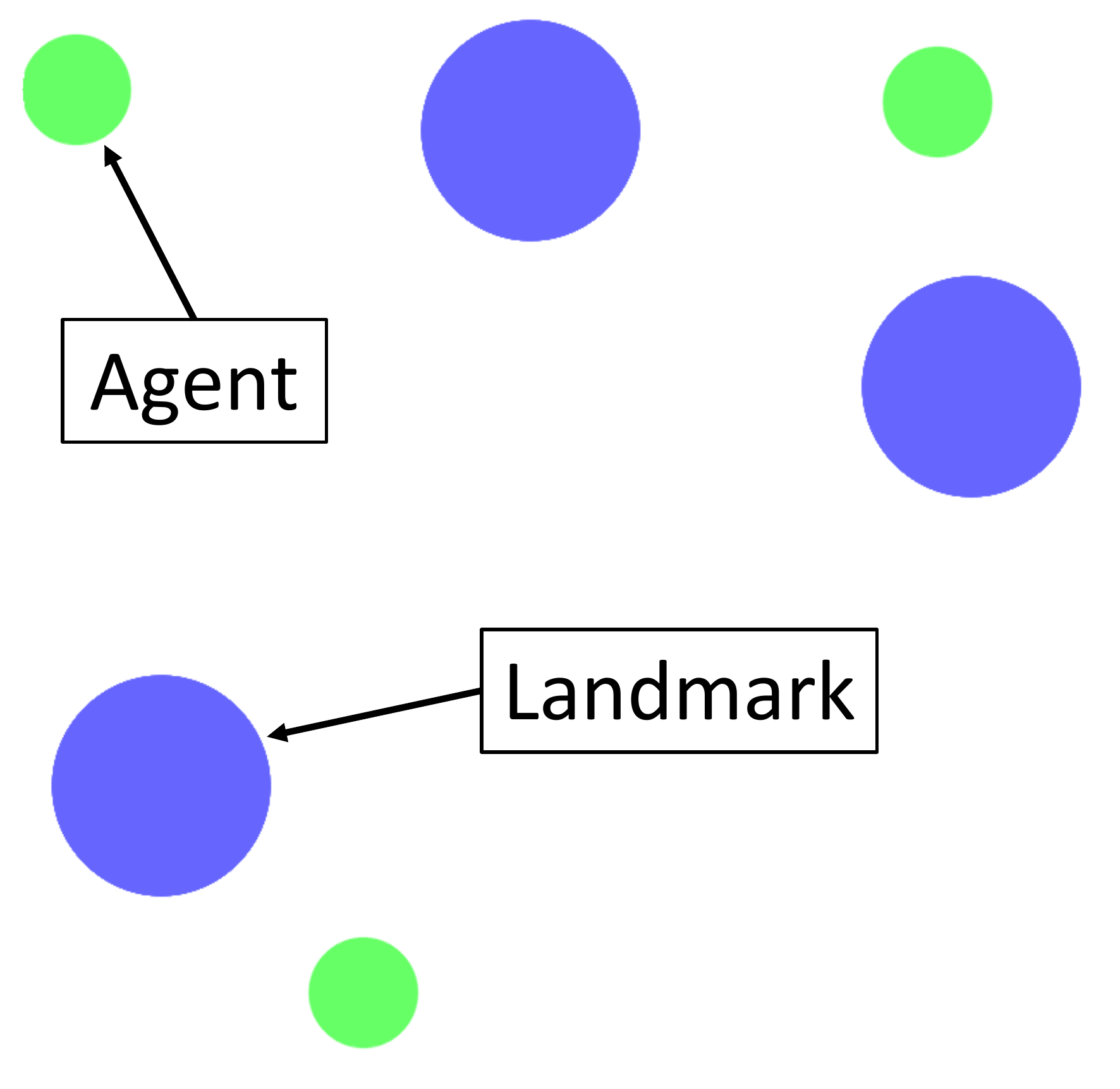} \\
     (a) & (b) & (c)
    \end{tabular}
    \caption{Considered environments: (a) Multi-walker, (b) Predator-prey, and (c) Cooperative navigation}
    \label{fig:environment}
\end{wrapfigure}

\textbf{Predator-prey} The predator-prey environment, which is a standard task for MARL, consists of $N$ predators and $M$ preys. We used a variant of the predator-prey environment into the continuous domain. The initial positions on the predators are randomly determined, and those of the preys are in the shape of a square lattice as shown in figure\ref{fig:environment} (b). The goal of the environment is to capture as many preys as possible during a given time $T$. A prey is captured when $C$ predators catch the prey simultaneously. The predators get team reward $R_1$ when they catch a prey. After all of the preys are captured and removed, we set the preys to respawn in the same position and double the team reward.
Thus, the different coordinated behavior is needed as $N$ and $C$ change. The observation of each agent consists of relative positions between agents and other agents and those between agents and the preys. Thus, each agent can access to all information of the environment state.
The action of each agent is two-dimensional physical action. We set $R_1=10$ and $T=100$. We simulated the environment with three cases: $(N=2, M=16, C=2$), $(N=3, M=16, C=1)$ and $(N=4, M=16, C=2)$.

\textbf{Cooperative navigation} Cooperative navigation, which was proposed in \cite{2017Lowe}, consists of $N$ agents and $L$ landmarks. The goal of this environment is to occupy all landmarks while avoiding collision with other agents. The agent receives shared reward $R_1$ which is the sum of the minimum distance of the landmarks from any agents, and the agents who collide each other receive negative reward $-R_2$. In addition, all agents receive $R_3$ if all landmarks are occupied. The observation of each agent consists of the locations of all other agents and landmarks, and action is two-dimensional physical action. We set $R_2=10$, $R_3=1$, and $T=50$. We simulated the environment in the cases of ($N=3$, $L=3$).

\newpage

\section*{Appendix E: Hyperparameter and Training Detail}

The hyperparameters for MA-AC, I-SAC, MA-SAC, MADDPG, and VM3-AC are summarized in Table \ref{table:app1}.

\begin{table}[h]
\caption{Hyperparameters of all algorithms}
\label{table:app1}
\vskip 0.15in
\begin{center}
\begin{small}
\begin{sc}
\begin{tabular}{lccccr}
\toprule
& MA-AC & I-SAC & MA-SAC & MADDPG & VM3-AC \\
\midrule
Replay buffer size    & $5\times 10^5$ & $5\times 10^5$ & $5\times 10^5$ & $5\times 10^5$ & $5\times 10^5$\\
Discount factor    & 0.99 & 0.99 & 0.99 & 0.99 & 0.99\\
Mini-batch size    & 128 & 128 & 128 & 128 & 128\\
Optimizer          & Adam & Adam& Adam& Adam& Adam \\
Learning rate   & 0.0003 & 0.0003 & 0.0003 & 0.0003 & 0.0003 \\
Target smoothing coefficient & 0.005 & 0.005 & 0.005 & 0.005 & 0.005 \\
Number of hidden layers (all networks) & 2 & 2 & 2 & 2 & 2 \\
Number of hidden units per layer & 128 & 128 & 128 & 128 & 128 \\
Activation function for hidden layer & ReLU & ReLU & ReLU & ReLU & ReLU \\
Activation function for final layer & Tanh & Tanh & Tanh & Tanh & Tanh \\
\bottomrule
\end{tabular}
\end{sc}
\end{small}
\end{center}
\vskip -0.1in
\end{table}

\begin{table}[h]
\caption{The temperature parameter $\beta$ for I-SAC, MA-SAC, and VM3-AC on the considered environments. Note that the temperature parameter $\beta$ in I-SAC and MA-SAC controls the relative importance between the reward and the entropy, whereas the temperature parameter $\beta$ in VM3-AC controls the relative importance between the reward and the mutual information. }
\label{table:ablation}
\vskip 0.15in
\begin{center}
\begin{small}
\begin{sc}
\begin{tabular}{lcccr}
\toprule
& I-SAC & MA-SAC & VM3-AC \\
\midrule
MW (N=3)    & 0.05 & 0.05 & 0.05\\
MW (N=4)    & 0.1 & 0.1 & 0.1\\
PP (N=2)    & 0.05 & 0.05 & 0.05\\
PP (N=3)    & 0.1 & 0.1 & 0.1\\
PP (N=4)    & 0.05 & 0.05 & 0.05\\
CN (N=3)    & 0.1 & 0.1 & 0.1\\
\bottomrule
\end{tabular}
\end{sc}
\end{small}
\end{center}
\vskip -0.1in
\end{table}

\begin{table}[h]
\caption{The dimension of the latent variable $z$ in VM3-AC}
\label{table:ablation}
\vskip 0.15in
\begin{center}
\begin{small}
\begin{sc}
\begin{tabular}{lcccr}
\toprule
& VM3-AC \\
\midrule
MW (N=3)    & 8 \\
MW (N=4)    & 8 \\
PP (N=2)    & 4\\
PP (N=3)    & 2 \\
PP (N=4)    & 4\\
CN (N=3)    & 8 \\
\bottomrule
\end{tabular}
\end{sc}
\end{small}
\end{center}
\vskip -0.1in
\end{table}

\newpage

\section*{Appendix F: Comparison against MAVEN}
\label{comparetomaven}

\begin{figure*}[h]
\begin{center}
\begin{tabular}{cc}
     % uncomment the next lines, and give the right ps files
     \includegraphics[width=0.35\textwidth]{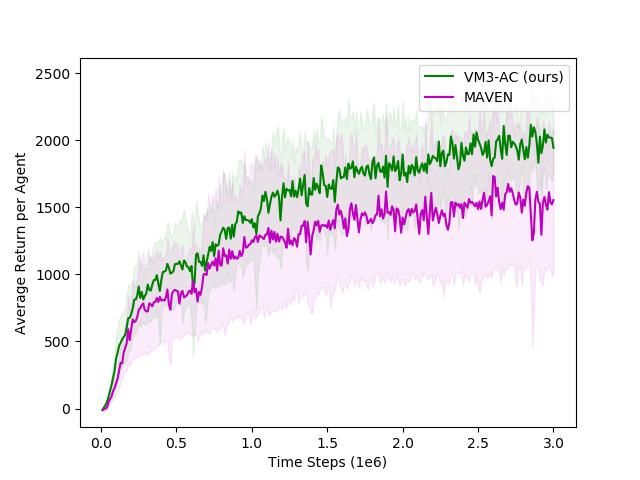} &
     \includegraphics[width=0.35\textwidth]{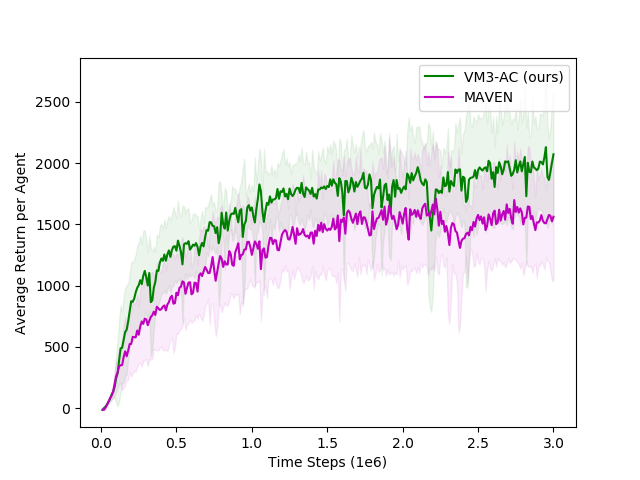} \\
     (a) Multi-walker (N=3) & (B) Multi-walker (N=4) \\
     \includegraphics[width=0.35\textwidth]{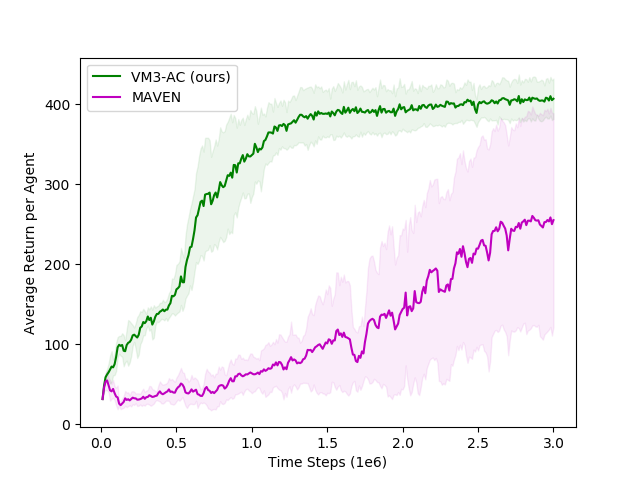} &
     \includegraphics[width=0.35\textwidth]{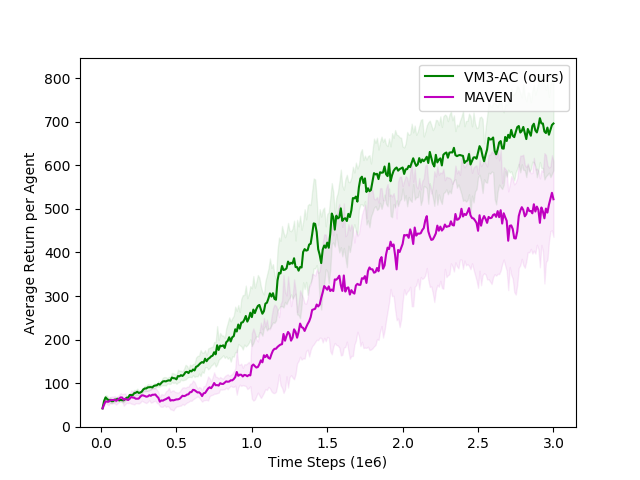} \\
     (C) Predator-prey (N=2) & (D) Predator-prey (N=3) \\
\end{tabular}
\caption{Comparison against MAVEN}
\label{fig:maven}
\end{center}
\end{figure*}

We compared the proposed VM3-AC algorithm with a very recent algorithm, MAVEN \cite{2019Mahajan2019}. Since MAVEN is based on the discrete action spaces, for comparison we applied the idea of MAVEN to actor-critic to devise a continuous action version. Then, we compared VM3-AC with this continuous-action version of MAVEN. The result is shown in  Fig. \ref{fig:maven}. It is seen that VM3-AC outperforms the continuous-action version of MAVEN. As seen in Figure \ref{fig:maven}, the performance gain of the proposed method over MAVEN is noticeable and that gain in the case of predator-prey with $N=2$ is drastic.

\end{document}